\documentclass[11pt]{article}
\usepackage{amssymb,amsmath} 
\usepackage{graphicx}

\def\={\equiv} 
\def\div{\nabla\cdot } 
\def\pl{\partial}

\newcommand{\bib}{\bibitem}

\newcommand{\nt}{\notag}
\newcommand{\ci}{\cite}
\newcommand{\lab}{\label}

\newcommand{\eq}{\eqref}

\newcommand{\bx}[1]{ \boxed{#1}}

\newcommand{\LB}{\left\lbrace}
\newcommand{\RB}{\right\rbrace}

\newcommand{\0}[1]{{(#1)}}
\newcommand{\1}[1]{{\hat #1}}

\newcommand{\3}[1]{{\boldsymbol #1}}

\newcommand{\bt}[1]{{\boldsymbol{\tilde #1}}}

\newcommand{\6}[1]{_{\scriptscriptstyle#1}}

\newcommand{\9}[1]{^{\,\scriptscriptstyle#1}}

\newcommand{\dtb}[1]{\dot{{\boldsymbol #1}}}
\newcommand{\ddtb}[1]{\ddot{{\boldsymbol #1}}}

\def\a{\alpha}

\def\d{\delta} 
\def\e{\varepsilon} 

\def\f{\phi}

\def\p{\pi}

\def\r{\rho}
 
\def\s{{\sigma}}

\def\D{\Delta} 
\def\F{\Phi}

\def\S{\Sigma} 

\def\Y{\Psi}

\newcommand{\hb}[1]{{\ \text{#1}\ }}

\newcommand{\db}{{\,{\rm d}\kern-.9ex {^-}}\!}
\newcommand{\dir}{{\pl\kern-1.2ex {/}}}

\newcommand{\curl}{\nabla\times}

\newcommand{\grad}{\nabla}

\newcommand{\im}{{\,\rm Im}\ }  
\newcommand{\imp}{\ \Rightarrow\ }

\newcommand{\ir}{\int_{-\infty}^\infty} 

\newcommand{\lra}{\leftrightarrow}

\newcommand{\Log}{{\rm Log\,}}

\newcommand{\plra}{\pl^{\kern-1.25ex^\lra}}
\newcommand{\qq}{\quad} 
\newcommand{\qqq}{\qquad} 
\newcommand{\re}{{\,\rm Re}\  }   
\newcommand{\rr}[1]{{{\mathbb R}^{#1}}}

\newcommand{\sr}{\sqrt}

\newcommand{\supp}{{\rm supp \,}}

\def\XXint#1#2#3{{\setbox0=\hbox{$#1{#2#3}{\int}$}
     \vcenter{\hbox{$#2#3$}}\kern-.5\wd0}}

\def\bib#1{\bibitem[#1]{#1}}
\begin{document}

\title{Making electromagnetic wavelets II:\\ Spheroidal shell antennas}

\author{Gerald Kaiser\\
Center for Signals and Waves\\
kaiser@wavelets.com  $\bullet$\ www.wavelets.com}


\maketitle

\begin{abstract}\noindent 
In the companion paper, a coherent charge-current distribution for radiating electromagnetic wavelets was constructed on an oblate spher\-oidal surface $\5S_\a$. Its main drawback was the necessity of including magnetic along with electric charges, making the sources impossible to realize.  Here we show how this difficulty can be overcome by using Hertz potentials to generate a charge-current distribution due solely to \sl bound electric charges. \rm  However, this distribution still appears difficult to realize because it consists of multiple surface layers on $\5S_\a$. We show how it can be replaced by a simple \sl volume \rm distribution on a spheroidal shell.  Our method generalizes the usual construction of equivalent \sl Huygens sources, \rm based on boundary conditions on an interface between electromagnetic media, by allowing the transition to be gradual without incurring addition complexity.
\end{abstract}

\section{Review of scalar wavelets}

We rely on the concepts in \ci{K4}, with some improvements in the notation. For further background on physical wavelets and complex-source pulsed beams, see  \ci{K94, HLK0, HF1}.  The complex distance from the imaginary source point $i\3a$ to the real observation point $\3r$ will be denoted by
\begin{align}\lab{sig}
\2r=\sr{(\3r-i\3a)\cdot(\3r-i\3a)}=p-iq
\end{align}
and the complex time by
\begin{align}\lab{tau}
\2t=t-ib.
\end{align}
The imaginary time $b$ plays the role of an overall \sl scale parameter, \rm similar to the scale of ordinary wavelets in one dimension, determining the \sl duration \rm of the pulsed-beam wavelets. The imaginary space vector $\3a$ similarly controls the \sl spatial \rm extent and orientation of the wavelets. The real and imaginary parts of $\2r$ satisfy the inequalities
\begin{gather}\lab{ineq}
|p|\le r,\   |q|\le a,\  
\hb{where}\  r=|\3r|,\   a=|\3a|.
\end{gather}
For fixed $\3a\ne\30$, the branch points of $\2r$ in $\rr3$ form the circle
\begin{align*}
\5C=\{\3r:\ \2r=0 \}=\{\3r:\ \3a\cdot\3r=0,\ r=a\}
\end{align*} 
and the `standard' branch of $\2r$, defined by $p\ge0$, has for its branch cut
the disk
\begin{align*}
\5D=\{\3r:\ p=0\}=\{\3r:\ \3a\cdot\3r=0,\ r\le a\}.
\end{align*} 
$\2r$ is real-analytic in $\rr3$ except for a jump discontinuity due to a sign reversal upon crossing $\5D$.
Every other branch satisfying the positivity condition
\begin{align*}
\3a\to\30\imp \2r\to r\ge 0
\end{align*}
can be obtained by continuously deforming $\5D$ to a membrane $\5B$ with the same boundary,
\begin{align*}
\pl\5B=\pl\5D=\5C. 
\end{align*} 
The associated branch of $\2r$ is defined by
\begin{align}\lab{rB}
\2r\6{\5B}=p\6{\5B}-iq\6{\5B}=\begin{cases}
\2r, & \3r\notin V\6{\5B}\\-\2r, &\3r\in V\6{\5B}
\end{cases}
\end{align}
where $V\6{\5B}$ is the compact volume swept out by deforming $\5D$ to $\5B$. It follows \ci{K4a} that $\2r\6{\5B}$ \sl  is real-analytic in $\rr3$ except for a sign reversal across $\5B$. \rm

The scalar wavelet\footnote{Strictly speaking, the term `wavelet' should be reserved for certain choices of $g$, as explained in \ci{K4}. The scalar wavelet of \sl order \rm $n$ is obtained with $\2g$ as the $n$-th derivative of the Cauchy kernel $C(\2t)=1/2\p i\2t$.
Also, note that we use units in which the propagation speed is $c=1$.}
with branch cut $\5B$ is defined by
\begin{align*}
\Y\6{\5B}=\frac{\2g(\2t-\2r\6{\5B})}{\2r\6{\5B}},
\end{align*}
where $\2g$ is the `analytic signal' associated to a driving signal $g\0t$ exciting the source by
\begin{align}\lab{g}
\2g(\2t)=\frac1{2\p i}\ir\frac{g(t')\,dt'}{\2t-t'},
\end{align}
which is indeed analytic in the complement of the support of $g$:
\begin{align}\lab{ganal}
\frac{\pl \2g(\2t)}{\pl\2t^*}=\frac12(\pl_t-i\pl_b)\2g(t-ib)=0\qq \forall \2t\notin \supp g\subset\4R.
\end{align}
The significance of the extension parameter $b$ can be understood by noting that the
real and imaginary parts of $\2g$ are
\begin{align}
g_b\0t&=\frac b{2\p }\ir\frac{g(t')\,dt'}{(t-t')^2+b^2}, &&
\1g_b\0t=\frac 1{2\p }\ir\frac{(t'-t) g(t')\,dt'}{(t-t')^2+b^2}. \lab{gb2}
\end{align}
$g_b$ is a smoothed version of $g$ with $b$ as the \sl scale \rm or \sl resolution parameter, \rm while $\1g_b$ is a smoothed version of the \sl Hilbert transform \rm of $g$,  again with $b$ as the scale parameter. Thus, \sl time variations of order less than $|b|$ are suppressed in $\2g(t-ib)$. \rm 

Due to its denominator, $\Y\6{\5B}$ is singular on $\5C$, where $\2r\6{\5B}=0$, and discontinuous in the interior of $\5B$, where $\2r\6{\5B}$ reverses sign.  
To avoid any further singularities, we want to ensure that the numerator $\2g(\2t-\2r\6{\5B})$ is analytic in all of $\rr3$, and for this it suffices to have its argument bounded away from the real axis by a positive distance.
Since
\begin{align}\lab{tr}
\2t-\2r\6{\5B}=t-p\6{\5B}-i(b-q\6{\5B})
\end{align}
and $|q\6{\5B}|=|q|\le a$, a necessary and sufficient condition is
\begin{align}\lab{ab}
a<|b|.
\end{align} 
This states that the imaginary space-time four-vector $(\3a, b)$ is \sl time-like, \rm belonging to the future cone of space-time if $a<b$ and the past cone if $b<-a$. The condition \eq{ab} will be assumed from now on, making $\Y\6{\5B}$ real-analytic in $\rr3-\5B$.

The \sl source \rm $\S\6{\5B}$ of $\Y\6{\5B}$ is  defined by applying the wave operator:
\begin{align}\lab{SB}
4\p\S\6{\5B}=\Box\Y\6{\5B} \qq\hb{where}\ \ \Box=\pl_t^2-\D.
\end{align}
$\S\6{\5B}$ can be easily shown to vanish wherever $\Y\6{\5B}$ is twice differentiable, hence
\begin{align}\lab{SB0}
\3r\notin\5B\imp \S\6{\5B}=0.
\end{align}
To characterize the source on $\5B$, we must apply $\Box$ in a \sl distributional \rm sense \ci{GS64}. Just as differentiating the Heaviside function gives the delta function, differentiating a discontinuous function like $\Y\6{\5B}$ in a distributional sense gives a \sl single layer \rm on the surface of discontinuity, represented by a delta function of a variable normal to that surface. Since $\S\6{\5B}$ is obtained by differentiating $\Y\6{\5B}$ \sl twice, \rm it will consist of a combination of \sl single and double layers \rm on $\5B$ \ci{K4, K4a}. Moreover, these layers diverge on the boundary $\5C=\pl\5B$ since $\Y\6{\5B}$ is singular there. This singularity will be tamed below by combining wavelets with different branch cuts.

\begin{figure}[ht]
\begin{center}
\includegraphics[width=4 in]{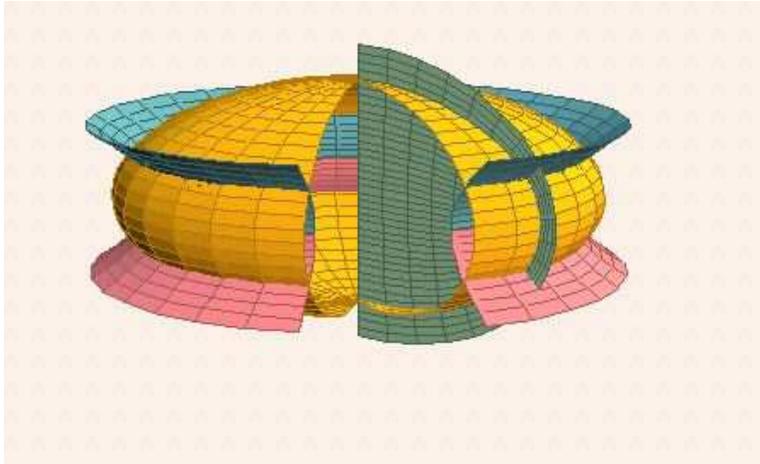}
\caption{$(p,q,\f)$ form an oblate spheroidal coordinate system.}
\label{FigOSCS}
\end{center}
\end{figure}

The variables $(p,q)$ defined by \eq{sig}, together with the azimuthal angle $\f$ about the $\3a$-axis, determine an \sl oblate spheroidal coordinate system \rm where the level surfaces of $p$ are the spheroids
\begin{align}\lab{Sp}
\5S_p:\qq\frac{x^2+y^2}{p^2+a^2}+\frac{z^2}{p^2}=1,\qqq p\ne 0
\end{align}
and the level surfaces of $q$ are the orthogonal hyperboloids  $\5H_q$. All these quadrics are \sl confocal, \rm having the circle $\5C$ as their common focal set. 
 This is depicted in Figure \ref{FigOSCS}.  As $p\to 0$, $\5S_p$ shrinks to a double cover of the disk $\5D$. 

When the \sl source \rm of $\Y\6{\5B}$ is computed, it will be singular on $\5C$ due to the singularity of $\Y\6{\5B}$ there. In the case of a real point source, this corresponds to the singularity of $\d(\3r)$ at $\3r=\30$. Recall that the latter can be regularized by replacing the origin with a sphere of small radius $r=\a$, whence $\d(\3r)$ is replaced by a uniform distribution on the spherical surface. The delta function can then be defined in terms of the limit $\a\to 0$. The equivalent procedure now is to replace the sphere by the oblate spheroid $\5S_\a$, which is defined by $p=\re\2r=\a>0$. But we can go a step further and represent $\5S_\a$ as a sum of two branch cuts, something that cannot be done for a real point source since the deformation of a point is still a point.
Thus, consider the  branch cut
\begin{align}\lab{Bp}
\5B\9+_\a=\5S_\a\9+\cup\5A_\a
\end{align}
consisting of union of the the upper hemispheroid 
\begin{align*}
\5S_\a\9+=\{\3r\in\5S_\a:\ z>0\}
\end{align*}
and the \sl apron \rm
\begin{align*}
\5A_\a=\{\3r:\ \3r\cdot \3a=0,\ a^2\le r^2\le a^2+\a^2\}
\end{align*}
connecting $\5S\9+_\e$ to the branch circle $\5C$. (The apron must be included so that  $\pl\5B\9+_\a=\5C$ as required.) 

\begin{figure}[ht]
\begin{center}
\includegraphics[width=4 in]{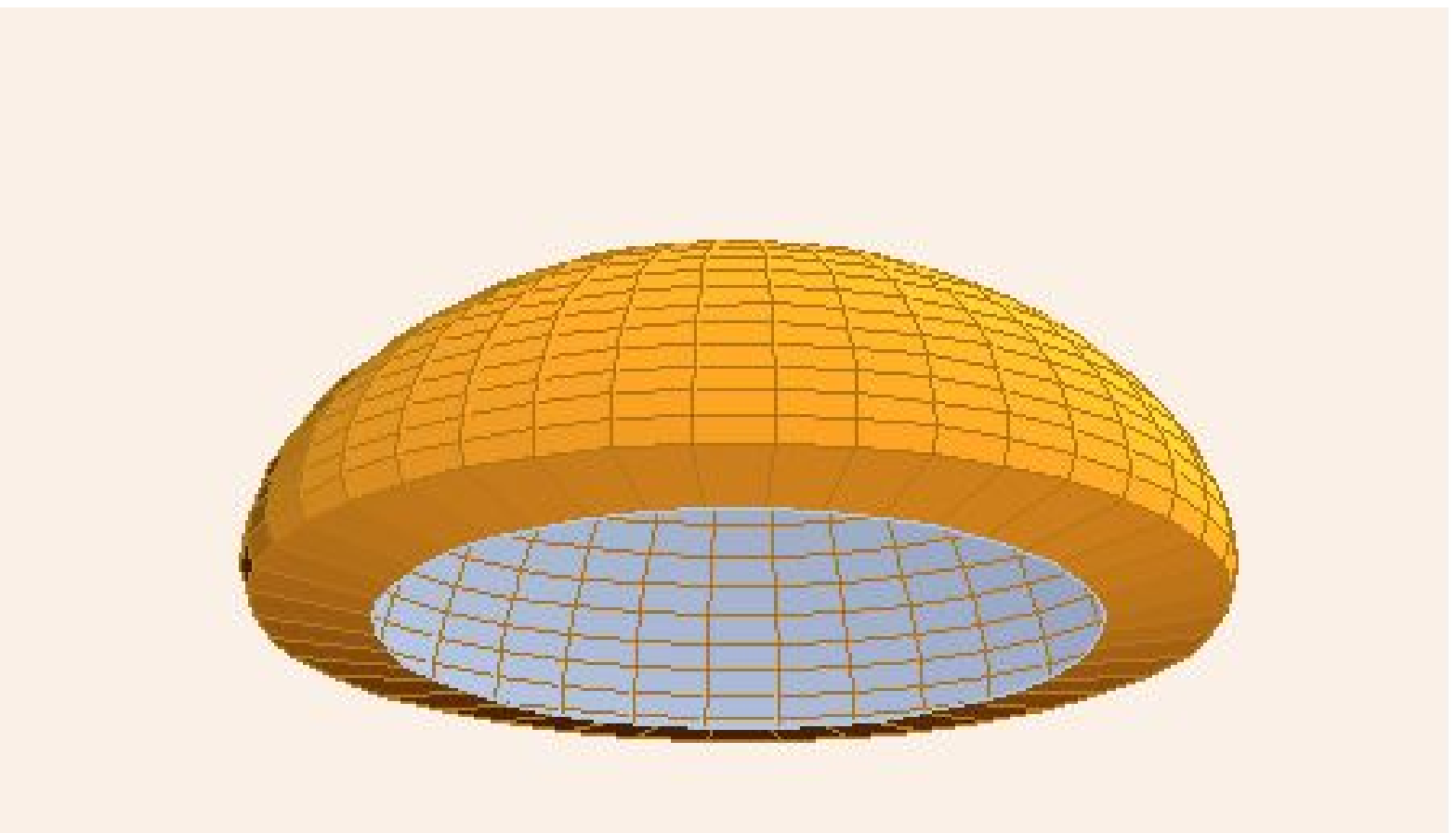}
\caption{The upper hemispheroidal branch cut $\5B\9+_\a$ with its apron $\5A_\a$.}
\label{FigOScut}
\end{center}
\end{figure}

Similarly, let
\begin{align}\lab{Bm}
\5B\9-_\a=\5S_\a\9-\cup\5A_\a
\end{align}
be the union of the lower hemispheroid
\begin{align*}
\5S_\a\9-=\{\3r\in\5S_\a:\ z<0\}
\end{align*}
with $\5A_\a$. 
For simplicity, denote the complex distance with branch cut $\5B\9\pm_\a$ by $\2r\6\pm$ instead of  $\2r\6{\5B_\a\9\pm}$ and the corresponding wavelet by $\Y\6\pm$. Let $V_\a\9\pm$ be the interiors of the upper and lower hemispheroids and  $V_\a$ be the interior of $\5S_\a$.  According to \eq{rB}, 
\begin{align*}
\3r\in V\9+_\a&\imp \Y\6+(\2r,\2t)=\Y(-\2r,\2t), \ \ 
 \Y\6-(\2r,\2t)=\Y(\2r,\2t)\\
\3r\in V\9-_\a&\imp \Y\6+(\2r,\2t)=\Y(\2r,\2t), \qq \ \,
\Y\6-(\2r,\2t)=\Y(-\2r,\2t)\\
\3r\notin V_\a&\imp \Y\6+(\2r,\2t)=\Y\6-(\2r,\2t)=\Y(\2r,\2t).
\end{align*}
Consider the \sl average \rm  of $\Y\6\pm$, 
\begin{align}\lab{Ya}
\Y_A(\2r,\2t)=\frac12\LB\Y\6-(\2r,\2t)+\Y\6+(\2r,\2t)\RB.
\end{align}
Then by the above, 
\begin{align}\lab{Ya1}
\Y_A(\2r,\2t)=\begin{cases} 
\Y_1(\2r,\2t), &\3r\in V_\a\\
\Y_2(\2r,\2t),& \3r\notin V_\a
\end{cases}
\end{align}
where the \sl internal \rm field $\Y_1$ and the \sl external \rm field $\Y_2$ are 
\begin{align}\lab{YIE}
\Y_1(\2r,\2t)=\frac12\LB\Y(\2r,\2t)+\Y(-\2r,\2t)\RB, \qqq
\Y_2(\2r,\2t)=\Y(\2r,\2t) 
\end{align}
and will, for later purposes, be regarded as functions on all of $\rr3$.
It follows directly from the definition \eq{g} of $\2g$ that
\begin{align}
\Y_1(\2r,\2t)&=
\frac1{4\p i\2r}\ir g(t')\, dt'\LB\frac1{\2t-t'-\2r}-\frac1{\2t-t'+\2r}\RB\nt\\
&=\frac1{2\p i}\ir\frac{g(t')\, dt'}{(\2t-t')^2-\2r^2}\,, \lab{Ya2}
\end{align}
which depends only on $\2r^2$ and is therefore \sl independent \rm of the choice of branch cut. Furthermore, since  \eq{ab} ensures that 
\begin{align*}
|\2t-t'\pm\2r|\ge |\im(\2t-t'\pm\2r)|= | b\pm q |\ge |b|-a>0,
\end{align*}
we have
\begin{align*}
|(\2t-t')^2-\2r^2|>(|b|-a)^2>0\ \forall \3r.
\end{align*}
This shows that $\Y_1$ is real-analytic in $\rr3$, at least if $g$ has compact support or decays sufficiently rapidly to ensure that the integral \eq{Ya2} converges. That is, by taking the average \eq{YIE} we have managed to cancel the singularities of $\Y(\pm\2r, \2t)$ on $\5C$ as well as their jump discontinuities across $\5D$, leaving a field which is analytic in all of $\rr3$ and hence \sl sourceless: \rm
\begin{align*}
\Box \Y_1(\2r,\2t)=0\ \forall (\3r, t)\in\rr4.
\end{align*}
The only region where $\Y_A$ fails to be analytic is therefore $\5S_\a$, where it is discontinuous by \eq{Ya1}. But even there, the irregularity is mild in the sense that the \sl jump discontinuity \rm
\begin{align} \lab{X}
\Y_J(\2r,\2t)&\=\Y_2(\2r,\2t)-\Y_1(\2r,\2t)
=\frac12\LB\Y(\2r,\2t)-\Y(-\2r,\2t)\RB
\end{align}
is \sl bounded \rm --- unlike that in any single branch $\Y\6{\5B}$, which is singular on $\5C$. Like $\Y_1$ and $\Y_2$, $\Y_J$ will be regarded as a field on all of $\rr3$, although for the present we need it only on $\5S_\a$. The  source $\S_A$ of $\Y_A$,  defined by
\begin{align}\lab{SA}
4\p\S_A=\Box\Y_A\,,
\end{align}
is therefore a distribution supported on $\5S_\a$. By the argument below  \eq{SB}, it
consists of a combination of single and double layers on $\5S_\a$, with the difference that these layers are now \sl bounded \rm since $\5S_\a$ avoids the singular circle $\5C$. Even so,  it is not clear that the double layer can be realized in practice. In the next section we replace $\Y_A$ by a \sl continuous \rm field, where the transition from $\Y_1$ to $\Y_2$ occurs gradually over a range of spheroids, whose source is supported on a spheroidal \sl shell \rm instead of a single spheroid. (This explains the reason for viewing $\Y_1, \Y_2$ and $\Y_J$ as fields on all of $\rr3$.) Such volume sources, and their electromagnetic counterparts considered in the following section, should be realizable.

\section{Spheroidal shell sources}

Let $ H$ be the Heaviside step function. Since $0\le p<\a$ in the interior of $\5S_\a$ and $p>\a$ in the exterior, we have
\begin{align}\lab{Ya3}
\Y_A= H(\a-p)\Y_1+ H(p-\a)\Y_2. 
\end{align}
It is natural to use the vector field orthogonal to $\5S_p$ given by \ci[Appendix]{K3} 
\begin{align}\lab{n}
\3n&\=\grad p=\frac{p\3r+q\3a}{p^2+q^2},
\end{align}
which is \sl unnormalized \rm with 
\begin{align}\lab{n2}
n^2\=|\3n|^2=\frac{p^2+a^2}{p^2+q^2}\ge 1\,,\qq \3n\cdot\grad q=0,\qq
\div\3n=\frac{2p}{p^2+q^2}.
\end{align}
We can compute the source $\S_A$ in \eq{SA} by using
\begin{align}\lab{gradQ}
\grad H(p-\a)=-\grad H(\a-p)= H'(p-\a)\3n=\d(p-\a)\3n.
\end{align}
Applying the wave operator gives $\S_A$ as a combination of terms involving $\d(p-\a)$, interpreted as  \sl single layers \rm on $\5S_\a$, and $\d'(p-\a)$, interpreted as \sl double layers. \rm As mentioned above, it is doubtful whether the double layer can be realized in practice, and this will get still worse in the electromagnetic case, where the currents involve one more derivative. Since we are interested in physically realizable sources, we now proceed to modify the above construction. The terms involving $\d(p-\a)$ and $\d'(p-\a)$ are unavoidable as long as the source is confined to the surface $\5S_\a$. To construct more realistic sources, we now choose a function $h$ that \sl approximates \rm the Heaviside function. A convenient example is
\begin{align}\lab{Q}
h\0p=\frac1\p \arg(-p+i\e)=\frac1\p\im\Log(-p+i\e),\qq \e>0
\end{align} 
which becomes the heaviside function in the limit as $\e\to 0$. 
Note that
\begin{align}\lab{sym}
 h'\0p=\frac\e{\p(p^2+\e^2)}
\end{align}
is indeed an approximation to the delta function. But the simple choice \eq{Q} has the drawback that $ h'$ is not compactly supported, so the resulting source, although extremely small outside a shell of thickness $\5O\0\e$, will not have strictly compact spatial support. To obtain a compactly supported source, we now assume that $ h$ has the following properties:
\begin{align}\lab{Q2}
h\0p+h(-p)=1\qq \hb{and} \qq
 h\0p=\begin{cases}0, &p\le-\e\\1,& p\ge \e\end{cases}
\end{align}
where $\e$ will be assumed fixed with
\begin{align}\lab{ep}
0<\e<\a,
\end{align}
so that $ h(p-\a)$ vanishes on the disk $\5D$  where $p=0$. Define the regularized version of $\Y_A$ by replacing $H$ by $h$ in \eq{Ya3},
\begin{align}\lab{YAe}
\Y^\e_A= h(\a-p)\Y_1 (\2r,\2t)+ h(p-\a)\Y_2 (\2r,\2t).
\end{align}
To simplify the equations, we use the abbreviations
\begin{align}\lab{Qpm}
h_1\0p= h(\a-p),\qq  h_2\0p= h(p-\a),
\end{align}
so that 
\begin{align}\lab{hsYs}
\Y_A^\e=h_1\Y_1+h_2\Y_2\=h_k\Y_k
\end{align}
where the Einstein convention of summing over repeated indices has been used.
The source $\S^\e_A$ of $\Y^\e_A$ is defined as usual by
\begin{align}\lab{S}
4\p \S^\e_A\=\Box \Y^\e_A.
\end{align}
To compute this, note that since \eq{Q2} implies $h'(-p)=h'\0p$, 
\begin{align}
\grad\Y^\e_A&=- h'(\a-p)\Y_1 \3n+ h'(p-\a)\Y_2 \3n+ h_k\grad\Y_k\nt\\
&= h'(p-\a)\Y_J\3n+ h_k\grad\Y_k \lab{gradF}
\end{align}
where $\Y_J$, defined as in \eq{X} and  given by
\begin{align}
\Y_J(\2r,\2t)&=
\frac1{4\p i\2r}\ir g(t')\, dt'\LB\frac1{\2t-t'-\2r}+\frac1{\2t-t'+\2r}\RB\nt\\
&=\frac1{2\p i\2r}\ir\frac{(\2t-t')g(t')\, dt'}{(\2t-t')^2-\2r^2}\,, \lab{YJ2}
\end{align}
no longer represents a jump discontinuity of the field since we are not confined to a single spheroid. By the same argument used to show that $\Y_1$ is real-analytic in $\rr3$, it follows that $\Y_J$ is real-analytic in $\rr3$ except for being discontinuous on $\5D$ and singular on $\5C$ due to the factor $1/\2r$.   Taking the divergence of \eq{gradF} gives
\begin{align*}
\D\Y^\e_A&= h''(p-\a) \Y_J n^2+2 h'(p-\a)\grad \Y_J\cdot\3n
+ h'(p-\a)\Y_J\div\3n
+ h_k\D\Y_k.
\end{align*}
Since $\Y_J(\2r,\2t)$ is complex-analytic in $\2r$ when $\3r\notin\5D$,
\begin{align*}
p>0\imp \grad \Y_J={\Y_J}'\,\grad \2r
\end{align*} 
where the prime denotes the complex derivative with respect to $\2r$,
\begin{align}\lab{YJr}
{\Y_J}'=\frac{\pl\Y_J}{\pl\2r}
=\frac12(\pl_p+i\pl_q)\Y_J=\frac12\LB\Y'(\2r,\2t)
+\Y'(-\2r,\2t)\RB.
\end{align}
By \eq{n2},  
\begin{align*}
\grad\2r\cdot\3n=(\grad p-i\grad q)\cdot\3n=n^2.
\end{align*}
Subtracting $\pl_t^2\Y^\e_A$ thus  gives
\begin{align*}
-4\p \S^\e_A= h''(p-\a) \Y_J n^2+2 h'(p-\a){\Y_J}'\, n^2
+ h'(p-\a)\Y_J\div\3n- h_k\Box\Y_k.
\end{align*}
But we have seen that $\Box\Y_1 $ vanishes identically and $\Box\Y_2 $ is supported on $\5D$, where $h_2=0$ by \eq{Q2} and \eq{ep}. 
Using \eq{n2} gives the regularized source
\begin{align}\lab{regS}\bx{
-4\p \S^\e_A= h''(p-\a) \Y_J n^2+\frac{2 h'(p-\a)}{p^2+q^2}
\LB (p^2+a^2){\Y_J}'\,+p \Y_J\RB}
\end{align}
supported on the spheroidal shell 
\begin{align}\lab{Shell}
\5S^\e_\a=\{\3r:\  \a-\e\le p\le \a+\e\}.
\end{align}
We emphasize that $\S^\e_A$ is a smooth volume source that depends only on the `jump field' $\Y_J$. Taking the limit $\e\to 0$ so that $h$ becomes the Heaviside function gives the source $\S_A$ consisting of single and double layers on $\5S_\a$.

\section{Maxwell's equations and Hertz potentials}

We work with the following complex combinations of electromagnetic fields:
\begin{align}
\3F&=\3D+i\3B\lab{F}\\
\3G&=\3E+i\3H=\3F-4\p\3P\lab{G}\\
\3P&=\3P_e+i\3P_m\lab{P}
\end{align}
where the units are Gaussian with $c=1$,  $\3P_e$ is the electric dipole density, and $\3P_m$ is the magnetic dipole density. Maxwell's equations take the form
\begin{align}\lab{Max} 
&\div\3F=4\p\r  && \dtb F+i\curl\3G=-4\p\3J
\end{align}
where $\dtb F=\pl_t\3F$. In the general case of \sl complex \rm charge and current densities
\begin{align*}
&\r=\r_e+i\r_m &&\3J=\3J_e+i\3J_m\,,
\end{align*}
equations \eq{Max} are equivalent to
\begin{align}\lab{Maxreal}
&\div\3D=4\p\r_e &&\dtb D-\curl\3H=-4\p\3J_e\\
&\div\3B=4\p\r_m &&\dtb B+\curl\3E=-4\p\3J_m\nt\,,
\end{align}
so the imaginary parts $(\r_m, \3J_m)$ represent the magnetic charge-current density.  Since magnetic monopoles are not observed, we must require
\begin{align}\lab{nomono}
\r_m=0 \hb{\  and\ } \3J_m=\30.
\end{align}
That is,  \eq{Max} are completely equivalent to the usual Maxwell equations if we add the requirement that $(\r,\3J)$ is real. We will consider solutions derived from a complex \sl Hertz potential \rm consisting of electric and magnetic Hertz vectors
\begin{align}\lab{Z}
\3Z=\3Z_e+i\3Z_m
\end{align}
whose source is the polarization,
\begin{align}\lab{P2}
\Box\3Z=4\p\3P.
\end{align}
(For this reason, $(\3Z_e, \3Z_m)$ are sometimes called \sl polarization potentials.\rm) 
The field $\3F$ is then given in terms of $\3Z$ by
\begin{align}\lab{FZ}
\3F=\curl\curl\3Z+i\curl\dtb Z,
\end{align}
and it follows from  \eq{G} and \eq{P2} that
\begin{align}
\3G&=\3F-\Box\3Z=\curl\curl\3Z+i\curl\dtb Z +\D\3Z-\ddtb Z\nt\\
&=\grad\div\3Z+i\curl\dtb Z-\ddtb Z. \lab{GZ}
\end{align}
The real form of equations \eq{P2},  \eq{FZ} and \eq{GZ} is \ci[pp 84--85]{BW99} 
\begin{align*}
&\Box \3Z_e=4\p\3P_e && \Box \3Z_m=4\p\3P_m\\
&\3E=\grad\div\3Z_e-\curl\dtb Z_m-\ddtb Z_e 
&&\3B=\curl\curl\3Z_m+\curl\dtb Z_e\\
&\3H=\grad\div\3Z_e+\curl\dtb Z_e-\ddtb Z_m
&&\3D=\curl\curl\3Z_e-\curl\dtb Z_m.
\end{align*} 
An inspection of the expressions for $\3E$ and $\3B$ reveals the meaning of the Hertz potentials as `superpotentials' from which the four-vector potential $(\F,\3A)$ can be derived by
\begin{align}\lab{4pot}
&\F=-\div\3Z_e  &&\3A=\curl\3Z_m+\dtb Z_e.
\end{align} 
In fact, these automatically satisfy the Lorenz condition 
\begin{align*}
\dot\F+\div\3A=0,
\end{align*}
and \sl every \rm four-vector potential satisfying it can be derived from Hertz potentials. The freedom to choose a  gauge for $(\F,\3A)$, including a non-Lorenz gauge, is part of a much greater gauge freedom in $(\3Z_e\,, \3Z_m)$  \ci{N55, BW99}.

According to \eq{FZ}, $\3F$ is a curl, so by \eq{Max} the free charge density vanishes: 
\begin{align}\lab{rho0}
\r=0.
\end{align}
Furthermore,  \eq{GZ} gives
\begin{align*}
\curl\3G=i\curl\curl\dtb Z-\curl\ddtb Z=i\dtb F,
\end{align*}
therefore by \eq{Max},  the free current density also vanishes: 
\begin{align}\lab{J0}
\3J=\30.
\end{align}
Maxwell's equations \eq{Maxreal}, written in terms of the \sl microscopic \rm fields 
$(\3E, \3B)$, now state that
\begin{align}
&\div\3B=0 &&\curl\3E+\dtb B=\30 \nt\\
&\div\3E=4\p\r_b &&\curl\3B-\dtb E=4\p\3J_b \lab{Maxreal2}
\end{align}
where 
\begin{align}\lab{Jb}
&\r_b=-\div\3P_e && \3J_b=\dtb P_e+\curl\3P_m
\end{align}
represent the \sl bound \rm charge and current densities generated by the variable polarizations $(\3P_e, \3P_m)$.  The fields derived from Hertz potentials as above are thus due entirely to bound sources.\footnote{Free charge-current densities can be added by using \sl stream potentials \rm \ci{N55}.}

\section{Spheroidal electromagnetic antennas}

In this section we construct electromagnetic wavelets from scalar wavelets
by turning $\Y$ into $\3Z$, then compute their charge-current densities.  It is essential that the polarization $\3P$ defined in \eq{P2} have compact spatial support, as it can otherwise not be realized. There are various ways to turn a scalar solution of the wave equation into a vector solution without increasing the support of its source distribution, the simplest being  
\begin{align}\lab{Z1}
\3Z=\3p\,\Y
\end{align}
where $\3p$ is a constant (possibly complex) vector. The polarization is then given by
\begin{align}\lab{P3}
4\p\3P=\3p\,\Box\Y=4\p\3p\,\S,
\end{align}
so $\3P$ and $\S$ have the same support.
Since $\S$ is a distribution consisting of single and double layers on $\5D$, so is $\3P$.  A similar construction applies to the different versions supported on the general branch cut $\5B$ and the spheroid $\5S_\a$. As explained below \eq{gradQ}, the layers on $\5B$ are singular on $\5C$ while those on $\5S_\a$ are bounded. Even so, the charge-current distributions \eq{Jb} require one further differentiation, hence they generate a still higher layer with coefficient $\d''(p-\a)$, and it is doubtful whether such distributions can be realized. For this reason we confine our analysis to volume sources on the spheroidal shell $\5S^\e_\a$ \eq{Shell}. Define the Hertz potential
\begin{align}\lab{2ZA}
\3Z^\e_A=\3p\,\Y^\e_A
\end{align}
with $\Y^\e_A$ as in \eq{YAe}, whose polarization density is
\begin{align}\lab{2PA}
\3P^\e_A=\3p\,\S_A^\e
\end{align}
with $\S^\e_A$ given by \eq{regS}. If we interpret $\S^\e_A$ as a scalar density, then \eq{2PA} suggests an interpretation of $\3p$ as a (complex) combination of \sl electric and magnetic dipole moments. \rm The charge and current densities on the shell, as given by \eq{Jb}, are
\begin{align} \lab{rb2}
\r_b=-\re\!\!\LB\div \3P^\e_A\RB=-\re\!\!\LB \3p\cdot\grad\S_\a^\e\RB
\end{align}
and
\begin{align}\lab{Jb2}
\3J_b&=\re\!\!\LB\dtb P^\e_A\RB+\im\!\!\LB\curl\3P^\e_A\RB
=\re\!\!\LB\3p\,\dot\S^\e_A\RB-\im\!\!\LB\3p\times\grad\S^\e_A\RB.
\end{align}
Outside the shell $\5S^\e_\a$ the potential $ \3Z^\e_A$ coincides with $\3Z=\3p\,\Y$, whose pulsed-beam field $\3F$ was computed in \ci{K4}.

\section{Extended Huygens sources}

The above suggests an generalization of Huygens sources \ci{HY99}, allowing equivalent sources to be represented on shells instead of surfaces surrounding a bounded source. We present this generalization and compare it to the usual method based on boundary conditions on an interface between electromagnetic media.  Let $p(\3r, t)$ be a differentiable function, which will be called a \sl zone function. \rm Fix two numbers $p_1<p_2$ and consider the time-dependent surfaces and volumes in $\rr3$ defined by
\begin{align*}
S_1\0t&=\{\3r: p(\3r, t)=p_1\},\qq S_2\0t=\{\3r: p(\3r, t)=p_2\}\\
V_1\0t&=\{\3r: p(\3r, t)< p_1\},\qq V_2\0t=\{\3r: p(\3r, t)> p_2\}.
\end{align*}
Given two electromagnetic fields $(\3F_1, \3G_1)$ and $(\3F_2, \3G_2)$,   with or without sources, we want to construct an \sl interpolated field \rm  $(\3F, \3G)$ so that
\begin{align}\lab{interpol}
\3F(\3r,t)&=\begin{cases} \3F_1(\3r,t), &\3r\in V_1\0t\\ 
\3F_2(\3r,t), &\3r\in V_2\0t\end{cases}\nt\\
\3G(\3r,t)&=\begin{cases} \3G_1(\3r,t), &\3r\in V_1\0t\\ 
\3G_2(\3r,t), &\3r\in V_2\0t.\end{cases}
\end{align}
Choose a differentiable function $h_2(\3r, t)$ such that
\begin{align}\lab{h2}
h_2(\3r,t)=\begin{cases} 0, & \3r\in V_1\0t\\ 1, &\3r\in V_2\0t \end{cases}
\end{align}
and let
\begin{align*}
h_1(\3r,t)=1-h_2(\3r, t).
\end{align*}
We define the interpolated field as
\begin{align}\lab{hsFs}
\3F(\3r,t)&=h_k(\3r,t)\3F_k(\3r,t)\nt\\ 
\3G(\3r,t)&=h_k(\3r,t)\3G_k(\3r,t)
\end{align}
where summations over $k=1, 2$ are implied, and the \sl jump field \rm
\begin{align}
\3F_J&=\3F_2-\3F_1=\3D_J+i\3B_J\nt\\
\3G_J&=\3G_2-\3G_1=\3E_J+i\3H_J. \lab{FGJ}
\end{align}
Then, according to \eq{Max}, the charge density of 
$(\3F, \3G)$ is
\begin{align}
4\p\r=\div\3F
=4\p h_k\r_k +\grad h_2 \cdot\3F_J \lab{r3}
\end{align}
where 
\begin{align}\lab{rIE}
4\p\r_k=\div\3F_k , \qq k=1,2
\end{align}
are the charge densities of the prescribed fields.  Thus, in addition to the interpolated charge density  
\begin{align}\lab{rI}
\r_I=h_k\r_k
\end{align}
we have a \sl transitional \rm  charge density given by
\begin{align}\lab{rT}
4\p\r_T=\grad h_2\cdot\3F_J
\end{align}
which depends only on the component of the jump field $\3F_J$ parallel to $\grad h_2$.  According to \eq{h2}, $\r_T$ vanishes outside the \sl  transition shell \rm
\begin{align}\lab{VT}
V_T\0t=\{\3r: p_1\le p(\3r,t)\le p_2\}.
\end{align}
Similarly, the current density is
\begin{align}\lab{J3}
4\p\3J&=-\dtb F-i\curl\3G=4\p h_k\3J_k-\dot h_2\3F_J-i\grad h_2\times\3G_J
\end{align}
where
\begin{align*}
4\p\3J_k=-\dtb F_k-i \curl\3G_k, \qq k=1,2
\end{align*}
are the current densities of the prescribed fields. Hence $\3J$
is the sum of the  interpolated current density 
\begin{align}\lab{JI}
\3J_I=h_k\3J_k
\end{align}
and a \sl transitional current density \rm on $V_T\0t$ given by
\begin{align}\lab{JT}
4\p\3J_T=-\dot h_2\3F_J-i\grad h_2\times\3G_J
\end{align}
which depends only on $\3F_J$ (if $h_2$ is time-dependent) and the component of $\3G_J$ orthogonal to $\grad h_2$. The electric and magnetic transitional sources are obtained by taking real and imaginary parts.  Assuming $h_2$ is \sl real, \rm this gives
\begin{align}
&4\p\r^e_T=\grad h_2\cdot\3D_J
&&4\p\3J^e_T=-\dot h_2\3D_J+\grad h_2\times\3H_J\nt\\
&4\p\r^m_T=\grad h_2\cdot\3B_J
&&4\p\3J^m_T=-\dot h_2\3B_J-\grad h_2\times\3E_J. \lab{JTem}
\end{align}
Letting $h_2$ be complex in $V_T\0t$ makes the transition shell a \sl chiral medium \rm mixing electric and magnetic fields. A further generalization is obtained by replacing $h_k$ with $3\times 3$ \sl matrices \rm (dyadics) $\4H_k$ satisfying
\begin{align}\lab{Hk}
&\4H_2(\3r,t)=\begin{cases} 0, & \3r\in V_1\0t\\ 
\4I, &\3r\in V_2\0t\end{cases}, \qqq \4H_1(\3r,t)=\4I-\4H_2(\3r,t)
\end{align}
where $\4I$ is the unit matrix. This makes the transition shell $V_T\0t$ a \sl non-isotropic \rm  medium as well as chiral if $\4H_k$ are complex. See \ci{LSTV94} for a treatment of chiral and non-isotropic media.

Choosing the zone function $p(\3r,t)$ time-dependent thus gives a simple formulation of the transition shell as a moving source, which could be useful in the analysis of radiation by moving objects. 

To see how all this relates to Huygens' principle, suppose we are only given a field $(\3F_2, \3G_2)$ whose charge-current density $(\r_2, \3J_2)$ is confined to $V_1\0t$, and want to find an \sl equivalent \rm charge-current density confined to $V_T\0t$ whose radiated field in $V_2\0t$ (but not necessarily elsewhere) is $(\3F_2, \3G_2)$. We are free to choose the field $(\3F_1, \3G_1)$ in any way that gives vanishing interpolated sources
\begin{align}\lab{nointerp}
\r_I=h_k\r_k=0, \qqq \3J_I=h_k\3J_k=\30,
\end{align}
since the sources of the interpolated field are then purely transitional and hence confined to $V_T\0t$ as desired. To satisfy \eq{nointerp}, it suffices to require that $(\r_1, \3J_1)$ be confined to $V_2\0t$. Thus, choosing any field 
$(\3F_1, \3G_1)$ with sources in $V_2\0t$ and any function $h_2$ satisfying \eq{h2},  an  equivalent charge-current density on $V_T\0t$ is given by \eq{rT} and \eq{JT}. As $p_1\to p_2$, $V_T\0t$ becomes $S_2$ and $(\r_T, \3J_T)$ become ordinary Huygens surface sources  \ci{HY99}.  

The freedom to choose $(\3F_1, \3G_1)$ (interpreted as the `interior field' if $V_1\0t$ is bounded) and $h_2$ is constrained by the requirement that the magnetic charge-current density must vanish, as detailed below.

Now suppose that $p=p(\3r)$ is time-independent, so $S_k$ and $V_k$ are fixed, and choose $h_k$ to be time-independent and real. As $p_1\to p_2$, assume that
\begin{align*}
\lim_{p_1\to p_2}\grad h_2(\3r) = \d(p(\3r)-p_2)\3n(\3r)
\end{align*}
where $\3n(\3r)$ is a normal vector field on $S_2$ pointing into $V_2$. Then \eq{rT} and \eq{JT} give 
\begin{align}\lab{sufsrc}
\r_T&\to\d(p(\3r)-p_2)\s\nt\\
\3J_T&\to\d(p(\3r)-p_2)\3K,
\end{align}
where
\begin{align}\lab{sufsrc2}
4\p\s=\3n\cdot\3F_J \hb{\ and\ }  4\p\3K=-i\3n\times\3G_J.
\end{align}
are the surface charge and current densities on $S_2$, whose
real and imaginary parts give the electric and magnetic surface sources:
\begin{align}
&4\p\s_e=\3n\cdot\3D_J 
&& 4\p\3K_e=\3n\times\3H_J\nt\\
&4\p\s_m=\3n\cdot\3B_J 
&& 4\p\3K_m=-\3n\times\3E_J. \lab{EMsurf}
\end{align}
Since magnetic monopoles are not observed, $\s_m$ and $\3K_m$ must vanish. When $p_1<p_2$, this may be accomplished if $h_2$ can be chosen so that
\begin{align}\lab{nomag}
\grad h_2\cdot\3B_J=0,\qqq \grad h_2\times \3E_J=\30,
\end{align}
which is possible\footnote{Letting $V_T$ be a nonisotropic medium by using $\4H_k$ \eq{Hk} makes it easier to enforce the absence of magnetic monopoles.}
if
\begin{align}\lab{EJBJ}
\3E_J\cdot\3B_J=0\qq\forall \3r\in V_T\0t.
\end{align} 
In the limit $p_1\to p_2$, \eq{nomag}  reduces \eq{EMsurf} to
\begin{align}
&\3n\cdot\3D_J =4\p\s_e
&& \3n\times\3H_J=4\p\3K_e\nt\\
&\3n\cdot\3B_J =0
&& \3n\times\3E_J=\30, \lab{EMsurf2}
\end{align}
which are the usual boundary conditions on an interface between two media.

Returning to the general time-dependent setting, consider now an alternative procedure of special interest here. Instead of interpolating two prescribed \sl fields \rm $(\3F_k, \3G_k)$, let us interpolate two Hertz potentials 
$\3Z_k$:
\begin{align}\lab{Z3}
\3Z=h_k\3Z_k.
\end{align}
As seen, this automatically results in vanishing `free' sources $\r=0,\ \3J=\30$. The polarization is found to be
\begin{align*}
4\p\3P&=\Box\3Z=4\p h_k\3P_k
+2\dot h_2\dtb Z_J-2(\grad h_2\cdot\grad)\3Z_J-\3Z_J\D h_2
\end{align*}
where $4\p\3P_k=\Box\3Z_k$ are the polarizations of the prescribed fields. If 
\begin{align*}
\supp \3P_1\subset V_2\hb{\ and\ } \supp \3P_2\subset V_1,
\end{align*}
then the interpolated polarization $h_k\3P_k$ vanishes and the polarization is purely transitional on $V_T\0t$:
\begin{align}\lab{P4}
4\p\3P&=2\dot h_2\dtb Z_J-2(\grad h_2\cdot\grad)\3Z_J-\3Z_J\D h_2.
\end{align}
This generalizes \eq{2PA}, and the bound charge-current densities derived from $\3P_e$ and $\3P_m$ via \eq{Jb} generalize \eq{rb2} and \eq{Jb2}.

\section{Conclusions}

We have improved on the computation of sources for electromagnetic wavelets given in \ci{K4} in two ways:  (a) The spheroidal surface $\5S_\a$ supporting the sources has been replaced by a spheroidal shell $\5S_\a^\e$ supporting  smooth \sl volume \rm sources. This eliminates the multiple layers on $\5S_\a$ which make the sources difficult if not impossible to realize. (b) By deriving the sources from Hertz potentials, we have eliminated the magnetic charge-current density, further facilitating their realizability. 

The problem with the magnetic sources in \ci{K4} can be better understood from the current perspective. Let the zone function be $p=\re\2r$ and $S_k, V_k$ be as above with $0<p_1<p_2$. Let the fields $(\3F_k, \3G_k)$ be derived from the Hertz potentials 
\begin{align*}
\3Z_k=\3p\Y_k, \qq k=1,2
\end{align*}
with $\Y_k$ given by \eq{YIE}. Recall that  $\Y_k$ are analytic in $(\2r,\2t)$  for $p>0$. Therefore the fields $(\3F_k, \3G_k)$ are analytic in  $(\bt r, \2t)$ for $\3r\notin\5D$, and 
\begin{align*}
\3P_k=\30,\qq   \3G_k=\3F_k=\3E_k+i\3B_k\qq \forall\3r\notin\5D.
\end{align*}
The jump fields
\begin{align*}
\3F_J=\3G_J=\3E_J+i\3B_J 
\end{align*}
are also analytic, as is their \sl polarization scalar \rm \ci{K3a}
\begin{align}\lab{FJ2}
\3F_J^2\=\3F_J\cdot\3F_J=\3E_J^2-\3B_J^2+2i\3E_J\cdot\3B_J.
\end{align}
The condition \eq{EJBJ} thus requires the imaginary part of an analytic function to vanish for $\3r\in V_T$, which implies that $\3F_J^2$ vanishes identically. 
Although every electromagnetic field must have $\3F_J^2\to 0$ in the far zone \ci{B15}, fields satisfying $\3F_J^2=0$ globally, called \sl null fields, \rm are 
rather degenerate. In particular, the electromagnetic wavelet fields and \sl not \rm null and hence cannot fulfill  \eq{EJBJ}. Instead, we have begun with an interpolated Hertz potential \eq{Z3} and derived bound sources in the transition shell, thus preserving analyticity  without invoking the existence of magnetic monopoles.

There is still an unsatisfactory  aspect to the polarization \eq{2PA} and charge-current density \eq{rb2}, \eq{Jb2}. Namely, they depend on the fixed vector  $\3p$ and thus do not conform to the spheroidal geometry. This suggests using methods of constructing $\3Z$ from $\Y$ other than \eq{Z1}. While \eq{Z1} is the complex version of \sl Whittaker's \rm potentials \ci{W4}, there are alternatives which do not require a fixed polarization vector, such as Debye potentials; see \ci{BD98} for example. Such alternatives will be considered in  future work.

\sl Note: \rm  After this paper was finished, I learned from Dr. Arthur Yaghjian that a similar formulation of the gradual transition between two electromagnetic media has been developed by Lindell, Tretyakov and Nikoskinen \ci{LTN0}. The present form is somewhat more general in the following respects. (a) The regions and surfaces are allowed to be time-dependent. (b) The derivation of generalized Huygens sources in  \ci{LTN0} makes the assumption that one of the prescribed fields vanishes, which we find to be unnecessary. This allows extended Huygens sources on $V_T$ with nonvanishing `internal' fields.

\section*{Acknowledgements}
It is a pleasure to thank Drs.~Richard Albanese, Grant Erdmann, Sherwood Samn and Arthur Yaghjian for helpful discussions.  I am also grateful to Dr.~Arje Nachman for his sustained support of my research, most recently through AFOSR Grant \#FA9550-04-1-0139.

\end{document}